\documentclass[prd,aps,floats,twocolumn,nofootinbib]{revtex4-1}
\usepackage{slashed}
\usepackage{mathtools}
\usepackage{amsfonts}
\usepackage{amssymb}
\usepackage{epsfig}
\usepackage{empheq}

\usepackage{bm}

\begin{document}
\newcommand*\widefbox[1]{\fbox{\hspace{2em}#1\hspace{2em}}}
\newcommand{\m}[1]{\mathcal{#1}}
\newcommand{\nn}{\nonumber}
\newcommand{\ph}{\phantom}
\newcommand{\eps}{\epsilon}
\newcommand{\be}{\begin{equation}}
\newcommand{\ee}{\end{equation}}
\newcommand{\bea}{\begin{eqnarray}}
\newcommand{\eea}{\end{eqnarray}}
\newtheorem{conj}{Conjecture}

\newcommand{\plk}{\mathfrak{h}}


\title{Dark matter and space-time symmetry restoration}

\date{}

\author{Jo\~{a}o Magueijo}
\email{j.magueijo@imperial.ac.uk}
\affiliation{Theoretical Physics Group, The Blackett Laboratory, Imperial College, Prince Consort Rd., London, SW7 2BZ, United Kingdom}

\begin{abstract}
We examine local physics in the presence of {\it global} variables: variables associated with the whole of the spacelike surfaces of a foliation. These could be the (pseudo-)constants of nature and their conjugate times, but our statements are more general. Interactions between the local and the global (for example, dependence of the local action on global times dual to constants) degrades full space-time diffeomorphism invariance down to spatial diffeomorphism invariance, and so an extra degree of freedom appears. When these presumably primordial global interactions switch off, the local action recovers full invariance and so the usual two gravitons, but a legacy matter component is left over, bearing the extra degree of freedom. Under the assumption that the preferred foliation is geodesic, this component behaves like dark matter, except that 3 of its 4 local degrees of freedom are frozen, forcing its rest frame to coincide with the preferred foliation. The non-frozen degree of freedom (the number density of the effective fluid) is the survivor of the extra ``graviton'' present in the initial theory, and keeps memory of all the past global interactions that took place in a given location in the preferred foliation. Such ``painted-on'' dark matter is best distinguished from the conventional one in situations where the preferred frame would be preposterous if all 4 degrees of freedom of dark matter were available. We provide one example: an outflowing halo of legacy matter with exact escape speed at each point and a very specific profile, surrounding a condensed structure made of normal matter.

\end{abstract}

\maketitle

\section{Introduction}
This work is both part of a project on variability of the laws of physics~\cite{evol,BHevol,MachianCDM}, and a repository of results likely applicable with some adaptation in the broader context of theories with a similar structure (e.g.~\cite{HL,shinji,Isham,brownkuchar,viqar,StueckelDM,Barvinsky,DBI}, reviewed further in our concluding Section). Suppose we foliate space-time into spatial leaves. Suppose some dynamical variables are functions of these leaves as a whole: we call them global variables. Finally, suppose that the local action depends parametrically on the global variables, but if that dependence were to disappear, the action would have 4D diffeomorphism invariance. In such theories, while global variables are interacting with the local physics (in a sense to be made precise in this paper), the local Hamiltonian constraint is lost, signaling the downgrading of full 4D diffeomorphism invariance to 3D spatial diffeomorphism invariance on each leaf. An extra local degree of freedom therefore appears (often blamed on an extra graviton mode). 

In this paper we show in great generality that this degree of freedom persists even after global interactions switch off, indeed it carries a memory of the integrated past effects of these interactions at a given point. However, this degree of freedom can then be neatly separated so that the local action regains full diffeomorphism invariance (and so, the standard two gravitons). The extra degree of freedom is absorbed by an effective matter component. If the preferred frame is geodesic (a major assumption to be made in this paper and dropped in its sequel~\cite{LIVCDM}), this component is equivalent to a pressureless fluid with 3 out of its 4 degrees of freedom frozen, pre-fixing its rest frame (identified with the fixed original foliation). A connection is thus made with dark matter, but this is ``matter without matter''. In addition, the presence of a preferred frame makes this effective dark matter  distinguishable in principle from conventional dark matter. 

We show in detail how this works without reference to any concrete theory, but a specific realization arises from taking for global variables the ``constants'' of Nature and their dual clocks~\cite{JoaoLetter,JoaoPaper}. A prototype for this construction is
the Henneaux-Teitelboim formulation~\cite{unimod} of unimodular gravity~\cite{unimod1,UnimodLee1,alan,daughton,sorkin1,sorkin2}, where the global variables are $\Lambda$ and 4-volume or unimodular time~\cite{Bombelli,UnimodLee2,JoaoLetter,JoaoPaper}, but other pairs can be chosen~\cite{pad,pad1,lomb,vikman,vikman1,vikaxion,JoaoLetter,JoaoPaper,evol}. If the clocks appear in the local action (i.e. if there is evolution) then there are global interactions and we fall within the remit of this paper.

\section{Preliminary definitions}\label{prelim}

We start with a few general definitions. We assume that the theory contains a preferred foliation $\Sigma_t$ (coordinatized by $(t,x)$). This is pre-fixed in a way that is to be further qualified later
(when we introduce a metric in the theory). 
Given $\Sigma_t$ we distinguish local and global variables (collectively represented by $\{q(x),p(x)\}$ and $\{\alpha,T_\alpha\}$) from the structure of the action:
\begin{eqnarray}
     S&=&\int dt\,  V_\infty \dot\alpha T_\alpha +S_0
     \nn\\
     &=&\int dt\,  \left[ V_\infty \dot\alpha T_\alpha
     + \int_{\Sigma_t} d^3x\, (\dot q(x) p(x) -{\cal H}_E)  \right]
\label{globdef}
\end{eqnarray}
where $V_\infty=\int_{\Sigma_t}  d^3 x$ is the volume of $\Sigma_t$, either finite or with a limiting procedure implied. The global variables are independent of $x$ on $\Sigma_t$, so they may be considered a property of the whole leaf. In contrast, each point on the leaf $x$ is associated with canonical pairs of local variables, $q(x)$ and $p(x)$.
The global variables, just as the local ones, are assumed to be {\it intensive} variables, which justifies the factor of $V_\infty$ in the first term of (\ref{globdef}).

The first term in~\eqref{globdef} contains the global variables already in their canonical Hamiltonian form. The global variables are assumed not to have a Hamiltonian of their own. The remaining terms, collected in $S_0$, make up the ``local'' action.
Upon a Legendre transformation, $S_0$ splits into the local variables' canonical terms and the extended Hamiltonian density, ${\cal H}_E$.  The non-vanishing Poisson brackets therefore are:
\begin{eqnarray}
    \{q(x),p(y)\}&=&\delta(x,y)\\
    \{\alpha,T_\alpha\}&=&\frac{1}{V_\infty}.
\end{eqnarray}
The local action $S_0$ is assumed to depend parametrically on the global variables (i.e. it does not depend on any derivatives of $\alpha$ or $T_\alpha$), so the same happens to ${\cal H}_E$. In what follows, the  dependence of ${\cal H}_E$ on the local variables can be generic, ranging from ultra-local (where ${\cal H}_E$ depends only on the local variables at that point) to local or perturbatively non-local (where it can depend on any order of gradients at that point). Thus, $S_0$ would perhaps be better dubbed ``non-global'', rather than ``local'' action, to distinguish it from the terms we wish to study. In contrast, the first term in \eqref{globdef} refers to variables that  exist globally, as a function $\Sigma_t$, not terms that depend non-locally on local variables.

To isolate effects due to global interactions it is useful to define the non-local Poisson bracket:
\begin{equation}
    \{f,g\}_{NL}=\frac{1}{V_\infty}\sum_{(\alpha ,T_\alpha)}
     \frac{\partial f}{\partial \alpha}\frac{\partial g}{\partial T_\alpha}- \frac{\partial f}{\partial T_\alpha}\frac{\partial g}{\partial \alpha}
\end{equation}
where $f$ and $g$ are general functions of phase space and the sum is over all global variable pairs.
Then, many expressions which in local theories are zero become proportional to the non-local Poisson bracket\footnote{The violations of energy conservation studied in~\cite{evol} (cf. Eq.82) are an example.}. We can also consider a corresponding ``local'' Poisson bracket, so that:
\begin{equation}
     \{f,g\}= \{f,g\}_{L}+\{f,g\}_{NL}.
\end{equation}

The equations of motion involve the total 
Hamiltonian $\mathbf{H}$, which is the space integral of ${\cal H}_E$:
\begin{align}
     {\bf H}&=\int_{\Sigma_t} d^3 x\, {\cal H}_E(x),
\end{align}
From this extensive quantity we can define an intensive total Hamiltonian density: 
\begin{align}
      {\cal H}_T&=\frac{{\mathbf H}}{V_\infty}
    =\frac{1}{V_\infty}\int_{\Sigma_t} d^3 x\, {\cal H}_E(x):
\end{align}
The Hamilton equations for the local variables are the usual:
\begin{eqnarray}
    \dot q(x)&=&\{q(x),{\bf H}\}=\frac{\delta  {\cal H}_E(x)}{\delta p(x)}
    \label{ham1loc}\\
     \dot p(x)&=&\{p(x),{\bf H}\}=-\frac{\delta {\cal H}_E(x)}{\delta q(x)}
\label{ham2loc}
\end{eqnarray}
computed at fixed $\alpha$ and $T_\alpha$ (just as one fixes all variables other than the pair involved). Hence, the evolution of local variables at a given point only depends on the functional derivatives of the Hamiltonian density ${\cal H}$ at that point.
This is to be contrasted with the equations of motion for the global variables. With our definition of ${\cal H}_T$ the factors of $V_\infty$ cancel out giving:
\begin{eqnarray}
    \dot\alpha&=&\{\alpha,\mathbf{H}\}=\frac{\partial {\cal H}_T}{\partial T_\alpha}
    \label{ham1totb}\\
    \dot T_\alpha&=&\{T_\alpha,\mathbf{H} \}=-
    \frac{\partial {\cal H}_T}{\partial \alpha}
    \label{ham2totb}
\end{eqnarray}
where only intensive variables appear.  The evolution of global variables therefore depends on the Hamiltonian density over the whole leaf. In this dependence, however, regions are weighted by their fractional volume: 
\begin{eqnarray}
    \dot\alpha&=&
    \int_{\Sigma_t} \frac{d^3 x}{V_\infty} \frac{\partial {\cal H}_E}{\partial T_\alpha}
    \label{ham1totb}\\
    \dot T_\alpha&=&-
    \int_{\Sigma_t} \frac{d^3 x}{V_\infty} \frac{\partial {\cal H}_E}{\partial \alpha},
    \label{ham2totb}
\end{eqnarray}
so that the larger the region, the more it weighs on the global variables. 
This is important in any setting, dubbed  Machian in~\cite{MachianCDM}, where a global presiding Universe acts as a ``reservoir''  fully determining the global variables, which can thus be seen as external to the equations of motion of any local system not part of the reservoir. 

\section{Symmetries and constraints}\label{symsconstr}
Our first task is to examine
the impact of global variables on the local space-time symmetries, the associated constraints and their algebra. We assume that the theory has a metric\footnote{So far we have only assumed a {\it spatial} foliation, so a conformal structure has been assumed.}, which upon an ADM split adapted to $\Sigma_t$ leads to lapse $N$, shift $N^i$, and induced or spatial metric $h_{ij}$. The extended Hamiltonian density takes the form:
\begin{equation}
    {\cal H}_E=N {\cal H} + N^i{\cal H}_i+...
\end{equation}
where ${\cal H}$ is the Hamiltonian density, ${\cal H}_i$ is the momentum density and the ellipsis denotes the possibility of other terms, to be explored elsewhere. 
We also assume that 
under the local Poisson bracket, $\cal H$ and ${\cal H}_i$ form the Dirac or hypersurface deformation algebra~\cite{Dirac,DiracCanadian,Thiemann,Bojo}, which here we present in its smeared version:
\begin{align}
    \{H_i(N^i), H_j(M^j)\}_L&= H_i([N,M]^i)\label{smearhihi}\\
    \{H_i(N^i), H(N)\}_L&= H(N^i\partial _iN)\label{smearhih0}\\
    \{ H(N), H(M)\}_L&= H_i(h^{ij}(N\partial_j M- M\partial_j N), \label{smearh0h0}
\end{align}
where $H(N)=\int d^3x N{\cal H}$ and $H_i(N^i)=\int d^3 x N^i {\cal H}_i$. The relation of this algebra to diffeomorphisms is well established. The statement that $S_0$ has 4D diffeomorphism invariance if global variables can be ignored is then the statement that the theory satisfies the Dirac algebra under the local part of the Poisson bracket, but not under the full bracket (including non-local variables). $S_0$ could be General Relativity.

The fact that global variables select a foliation 
should put us on notice: we should not start by assuming more than spatial diffeomorphism invariance on leaves $\Sigma_t$, leaving the status of full space-time diffeomorphisms on hold. (Even the assumption of spatial diffeomorphism invariance may have to be questioned later, but let us see how far it can take us.) Note that regarding spatial diffeomorphisms, $q$, $\alpha$, $N$ and $N^i$ are tensors, and that $p$, $T_\alpha$, ${\cal H}$ and ${\cal H}_i$ are tensor densities of weight 1, balancing out the weight of $d^3x$ or $V_\infty$, so that the theory is spatially diffeomorphism invariant. 

Since we assume spatial diffeomorphism invariance, $N^i$ should be allowed to be a general function $N^i(t,x)$, so that its variation implies the local momentum constraint:
\begin{equation}\label{diffconst}
    {\cal H}_i(x)\approx 0. 
\end{equation} 
Its preservation requires the vanishing of 3 terms:
\begin{align}
    \dot {\cal H}_i=\{ {\cal H}_i,\mathbf{H}\}
    =& \{ {\cal H}_i,\mathbf{H}_0\}_{L} +\{ {\cal H}_i,\mathbf{P}\}_{L} + \{ {\cal H}_i,\mathbf{H}\}_{NL}\nn
\end{align}
where we split $\mathbf{H}$ as: 
\begin{align}
  {\bf H}_0&\equiv H(N)=\int_{\Sigma_t} d^3 x\, N {\cal H}(x)\\
  \mathbf{P}&\equiv H_i(N^i)=\int_{\Sigma_t} d^3 x\, N^i {\cal H}_i(x). 
\end{align}
The first term  becomes\footnote{In the various calculations that follow there will be integrations by parts. Strictly speaking we either have no spatial boundary, or boundary terms would have to be added.}:
\begin{equation}\label{HorN}
    \{ {\cal H}_i(x),\mathbf{H}_0\}_L={\cal H}(x)\frac{\partial N}{\partial x^i},
\end{equation}
(using \eqref{smearhih0}). 
The second term is zero on-shell (using  \eqref{smearhihi} and \eqref{diffconst}). The third, non-local, term implies:
\begin{equation}
     \{ {\cal H}_i,\mathbf{H}\}_{NL}=\frac{\partial {\cal H}_i}{\partial \alpha}\dot \alpha+ \frac{\partial {\cal H}_i}{\partial T_\alpha}\dot T_\alpha.
\end{equation}
due to \eqref{ham1totb} and \eqref{ham2totb}, and must be zero separately. Except in trivial situations this places constraints on the dependence of ${\cal H}_i$ on the global variables. 
If the dependence on global variables can be factorized as: 
\begin{equation}\label{condHi}
    {\cal H}_i =F(\alpha,T_\alpha)\bar {\cal H}_i
\end{equation}
with $\bar {\cal H}_i$ independent of global variables, we have for any function $f$ of phase space:
\begin{equation}\label{NLHi}
    \{f,{\cal H}_i\}_{NL}\approx 0.
\end{equation}
We will assume this from now onwards. This assumption amounts to the statement that the effect of global variables on the momentum constraint may be absorbed into a redefinition of $N^i$. More trivially, the momentum constraint may not depend on any global variables (just as it does not depend on derivatives of the metric in most theories).

We are thus left with condition \eqref{HorN}, which presents us with two possibilities for consistency of spatial diffeomorphism invariance on $\Sigma_t$, and the fate of the Hamiltonian constraint. Either we also have a local Hamiltonian constraint, ${\cal H}(x)=0$, in which case $N$ can be general (and if so, indeed this implies ${\cal H}=0$); or we do not, in which case $N$ {\it cannot} depend on space.
To investigate where we stand regarding ${\cal H}=0$ we evaluate its would-be secondary constraint:
\begin{align}
    \dot {\cal H}=\{ {\cal H},\mathbf{H}\}
    =& \{ {\cal H},\mathbf{H}_0\}_{L} +\{ {\cal H},\mathbf{P}\}_{L} + \{ {\cal H},\mathbf{H}\}_{NL}\nn.
\end{align}
The first terms vanishes on-shell (using  \eqref{smearh0h0} and \eqref{diffconst}). 
The second term gives: 
\begin{equation}
  \{{\cal H}(x),\mathbf{P}\}_L\approx D_i(N^i{\cal H})
\end{equation}
using \eqref{smearhih0}. Hence we get a master equation for the variation of the Hamiltonian, assuming the diffeomorphism constraint: 
\begin{equation}\label{dotHeq}
\boxed{
    \dot {\cal H}=D_i(N^i{\cal H})+ \{{\cal H},\mathbf{H}\}_{NL}.}
\end{equation}
Regarding the last term, 
notice that (\ref{NLHi}) implies 
$
\{{\cal H},\mathbf{P}\}_{NL}=0 $,
but we are still left with:
\begin{equation}
    \{{\cal H},\mathbf{H}\}_{NL}
    = \{{\cal H},\mathbf{H}_0\}_{NL}=\frac{\partial {\cal H}}{\partial \alpha}\dot \alpha + \frac{\partial {\cal H}}{\partial T_\alpha}\dot T_\alpha.
\end{equation}
The vanishing of this expression is therefore a sine qua non for a possible ${\cal H}=0$. {\it We thus use its non-vanishing as a definition for global interactions being switched on.} 

If global interactions are switched on:
\begin{equation}\label{globalONOFF}
     \{{\cal H},\mathbf{H}\}_{NL}\neq 0\implies \dot {\cal H}\neq 0 \implies {\cal H}\neq 0
\end{equation}
so that the consistency of spatial diffeomorphisms (cf. \eqref{HorN}) forces 
\begin{equation}\label{Njustoft}
    \partial_i N=0\implies N=N(t)
\end{equation}
and variation with respect to $N$ gives only a {\it global}, rather than a local,  Hamiltonian constraint:
\begin{equation}\label{globalHconst}
    \mathbf{H}_0=0.
\end{equation}
This global condition does not generate any new secondary constraints, since:
\begin{align}
    \dot {\mathbf{H}}_0&=\{ \mathbf{H}_0, \mathbf{H}\}=\{ \mathbf{H}_0, \mathbf{P}\}=0
\end{align}
in view of \eqref{NLHi}. We are left with first class constraints corresponding to spatial diffeomorphisms and a global time reparameterization invariance of leaves $\Sigma_t$. Their associated constraints are the usual 4 global constraints, but only 3 local ones. Hence we gain a local degree of freedom while global interactions are switched on. 

We will not dwell upon its significance while global interactions are switched on, but rather investigate whether full diffeomorphism invariance could ever be restored in such theories once global interactions switch off
(as defined by the negative of \eqref{globalONOFF}) and if so, where the extra degree of freedom goes under such circumstances. 

\section{Symmetry restoration post-global interactions}\label{Sec:restor}

Equation \eqref{dotHeq} is very interesting. Recall that $\cal H$ is a (3D) spatial density of weight 1, so the first term on its right hand side is the spatial Lie derivative on $\Sigma_t$:
\begin{equation}
    \pounds_{{N^i}}{\cal H} =\partial_i(N^i{\cal H})= D_i(N^i{\cal H})
\end{equation}
where $D_i$ is the spatial covariant derivative. 
If ${\cal H}\neq 0$ and global interactions are switched off, then the only time variations in $\cal H$ are those due to the spatial diffeomorphisms associated with $N^i$ 
(this generalizes the result found in~\cite{MachianCDM}: if $N^i=0$ the legacy Hamiltonian is a constant in time). 

We can recast this fact as the vanishing of a {\it total time derivative} defined as the 4D Lie derivative with respect to the normal $n^\mu$ to $\Sigma_t$ and the congruence $\cal C$ of its integral lines. With the ADM convention:
\begin{align}
     n_\mu&=(-N,0)\label{n_mu}\\
     n^\mu&=\left(\frac{1}{N},\frac{-N^i}{N}\right)\label{n^mu}
\end{align}
(where both $N$ and $N^i$ can a priori depend on $t$ and $x$)
and noticing that $N{\cal H}$ is a 4D density, 
we can work out the identity (true for any 4D density of weight 1 built from $N$ and a 3D density):
\begin{align}
      \pounds_{n^\mu}(N{\cal H} ) = & 
n^\mu\partial_\mu (N{\cal H} )+(\partial_\mu n^\mu ) N{\cal H} \nn\\
      =&  \dot {\cal H}-\partial_i(N^i{\cal H})=
    \dot {\cal H}-D_i(N^i{\cal H})\label{4DLie}
\end{align}
(in the algebraic manipulations we did {\it not} use $\partial_i N=0$).  Then \eqref{dotHeq} becomes:
\begin{equation}
\boxed{
     \frac{D{\cal H}}{dt}\equiv  
     \pounds_{n^\mu}(N{\cal H} )=
     \{{\cal H},\mathbf{H}\}_{NL}
   =\frac{\partial {\cal H}}{\partial \alpha}\dot \alpha + \frac{\partial {\cal H}}{\partial T_\alpha}\dot T_\alpha.}
\end{equation}
When global interactions are switched on the Hamiltonian is driven by a source, leading to an expression which can be formally written as
\begin{equation}
    {\cal H}(t,x)=\int_{\cal C} \left(\frac{\partial {\cal H}}{\partial \alpha}\dot \alpha +\frac{\partial {\cal H}}{\partial T_\alpha}\dot T_\alpha\right)
\end{equation}
that is the line integral of the global interactions along the congruence $\cal C$ tangent to $n^\mu$ (this is just the integral in time if $N^i=0$). When the interactions are switched off one is left with a generically non-zero, space-dependent integration constant which is a memory of all past interactions.  The free evolution of this non-zero Hamiltonian is equivalent to Lie dragging $N{\cal H}$ along the vector $n^\mu$ normal to $\Sigma_t$. 

It would therefore seem that we are left with a permanent blemish: full diffeomorphism invariance once broken is gone forever, even after global interactions switch off. 
This does not need to be the case. 
Let us {\it assume} that the base theory $S_0$ in the absence of global interactions (past, present and future) has full diffeomorphism invariance\footnote{This is to be contrasted with the equivalent problem in Horava-Lifshitz theory~\cite{HL,shinji}, subject to global interactions~\cite{HLPaolo} or otherwise. In that case there is nothing to restore, $N(t)$ can never be replaced by $N(t,x)$, and the Hamiltonian constraint remains essentially global.}. After global interactions switch off, defined as: 
\begin{equation}\label{switchoff}
     \{{\cal H},\mathbf{H}\}_{NL}=0
\end{equation}
(for example due to $\partial {\cal H}/\partial T_\alpha=0 $), we can rewrite the Hamiltonian as:
\begin{equation}\label{barH}
    \bar {\cal H}={\cal H}(q(x),p(x);\alpha) +m\approx 0
\end{equation}
with $m(x,t)$ a 3D density evolving according to:
\begin{equation}\label{LieDrag}
    \frac{D m}{dt}=
    \dot m-\partial_i(N^i m)=0
\end{equation}
and initial conditions set at the $\Sigma_{t_0}$ leaf when global interactions were switched off. These initial conditions contain the integrated effect of all previous global interactions and can be formally written as:
\begin{equation}
    m(t_0,x)=-\int_{\Sigma_{-\infty}}^{\Sigma_{t_0}} \left(\frac{\partial {\cal H}}{\partial \alpha}\dot \alpha+ \frac{\partial {\cal H}}{\partial T_\alpha}\dot T_\alpha\right)
\end{equation}
over the line $\cal C$ linking the two $\Sigma_t$ surfaces.  The newly found local Hamiltonian constraint \eqref{barH} can be enforced by allowing $N$ to depend on space:
\begin{equation}\label{Nofx}
    N(t)\rightarrow N(t,x)
\end{equation}
in the corresponding action $\bar S_0$ built from $\bar {\cal H}$. In turn, as we will see, the evolution \eqref{LieDrag} can be enforced by giving $m$ a canonical partner $\phi$ and adding to the momentum constraint a new term:
\begin{equation}
    \bar {\cal H}_i ={\cal H}_i + m\, \partial_i \phi
\end{equation}
with $\partial_i\phi=0$ on $\Sigma_t$ 
enforced by a new constraint  $\lambda^i\partial_i\phi$.
Then $\bar S_0$, as written in coordinates adapted to foliation $\Sigma_t$, reads:
\begin{align}
     \bar S_0=&
     \int dt\int_{\Sigma_t} d^3x\, (\dot q(x) p (x)+\dot \phi(x) m(x)\nn\\
     &-N(t,x) \bar {\cal H} -N^i(t,x)\bar{\cal H}_i +\lambda^i \partial_i\phi   ),
\end{align}
which splits neatly into:
\begin{align}\label{barS0}
     \bar S_0=&S_0+S_m.
\end{align}
Here
\begin{align}
    S_0=& \int dt\int_{\Sigma_t} d^3x\, (\dot q(x) p (x)-N(t,x) {\cal H} -N^i(t,x) {\cal H}_i) 
\end{align}
is nothing but the local action we started from, but with $N=N(t,x)$, that is with full diffeomorphism invariance restored. The price to pay for this is the new term:
\begin{align}
    S_m=&  \int dt\int_{\Sigma_t} d^3x\, (\dot \phi(x) m(x)
     -N(t,x)m\nn\\
     &-N^i(t,x)m\partial_i\phi -\lambda^i\partial_i\phi)\label{SmprefS}
\end{align}
to be examined for the rest of the paper. 

The blemish in diffeomorphism invariance is therefore contained. 
$S_0$ now has all the properties of the base theory as if no global interactions ever existed. If $S_m$ can be ignored, then 4D diffeomorphism invariance is restored. This is true even ignoring $S_m$ just conceptually: when we evaluate the number of degrees of freedom of gravity, we mean ``in vacuum" in the sense of ignoring the matter degrees of freedom. By separating the legacy effects of past global interactions, we can call them a matter component. Then the gravity action $S_G$ contained in $S_0$ has 4 first class constraints, and so 2 gravitons as usual. $S_m$ adds degrees of freedom, just like any other matter component adds degrees of freedom. We will now examine the properties of the extra degree of freedom transferred to $S_m$.

\section{An equivalence principle for foliations}

Focusing on $S_m$ and the consistency of the constraints it generates we find that a central issue is how the supposedly ``pre-fixed'' $\Sigma_t$ interacts with the metric, and specifically the lapse function, once $N$ is allowed to have spatial dependence on $\Sigma_t$ {\it in the variational problem}. The question is whether $N$ can be made to remain solely a function of $t$ {\it on-shell}, even when we allow it to vary in space in the variational problem, cf.~\eqref{Nofx}. This sets up a major fork in the formalism, and is related to the geodesic nature of the preferred $\Sigma_t$.

In this paper we impose a ``weak equivalence principle'' for preferred foliations, stating that the foliations remain geodesic even after the lapse function's local degrees of freedom are released. This can be phrased as:
\begin{equation}
    a_\mu =n^\nu\nabla_\nu n_\mu=0
\end{equation}
and is equivalent to the statement that the (normalized) normal $n^\mu$ is proportional to a gradient:
\begin{equation}
    n_\mu=\partial_\mu T
\end{equation}
for some scalar $T$ (which could be any global variable $T_\alpha$ which has dropped out of the dynamics). In turn this is equivalent to $N$ being only a function of $t$ on the preferred $\Sigma_t$. More directly we could use (see~\cite{ADMReview} for an excellent primer):
\begin{equation}
    a_\mu=h_\mu^{\;\nu} \nabla_\nu \ln N
\end{equation}
with $h_{\mu\nu}=g_{\mu\nu}+n_\mu n_\nu$ the  projector orthogonal to $n_\mu$ (and intrinsic metric $h_{ij}$).

Thus, whilst $\Sigma_t$ had to be geodesic under the action of spatially diffeomorphism invariant global interactions (cf. \eqref{Njustoft}), we are now postulating that this remains true after they switch off. If $\Sigma_t$ is geodesic, then our coordinates may be chosen so that $N=N(t)$, that is, although we have allowed $N$ to depend on space in the variations leading to the Hamiltonian constraint (cf.~\eqref{Nofx}), we can choose coordinates such that on-shell $N=N(t)$. (This is not unusual. In a related example which we will find in Section~\ref{Sec:test}, the LTB metric has $N^i=0$ and yet it has a non trivial momentum constraint, resulting from varying $N^i(t,x)$.)

It is not difficult to see that if we drop this assumption the rest of the story is very different. If $\Sigma_t$ is non-geodesic, after global interactions switch off we have on $\Sigma_t$:
\begin{align}
   \dot {\cal H}=\{{\cal H},\mathbf{H}\}_L&=\partial_i(N^i {\cal H})+\partial_i ({\cal H}^i N)+ {\cal H}^i \partial_i N \label{dotHH}\\
        \dot {\cal H}_i=\{{\cal H}_i,\mathbf{H}\}_L&= {\cal H}\, \partial_i N +D_j(N^j {\cal H}_i)  + {\cal H}_jD_i N^j\nn\\
        &= {\cal H}\, \partial_i N +\partial_j(N^j {\cal H}_i)  + {\cal H}_j\partial _i N^j
        \label{dotHHi}
\end{align}
(where ${\cal H}^i=h^{ij}{\cal H}_j$ 
and $D_i$ is the covariant derivative compatible with $h_{ij}$\footnote{Note that, as was done explicitly for the last two terms as a block, all $\partial_i$ can be replaced by $D_i$.})
which can be derived using \eqref{smearhihi}-\eqref{smearh0h0}. These evolution equations are inherited by ${\cal H}_m$ and ${\cal H}^m_i$ upon rewriting:
\begin{align}
     \bar {\cal H}&={\cal H}(q(x),p(x);\alpha) +{\cal H}^m\approx 0\\
     \bar {\cal H}_i&={\cal H}_i(q(x),p(x);\alpha) +{\cal H}_i^m\approx 0.
\end{align}
Requiring that ${\cal H}^m_i\approx 0$ as we did in the last section therefore requires $N=N(t)$ on-shell. Hence the geodesic assumption for the preferred frame, the statement that there are coordinates for which $N=N(t)$, and the constraint that an initially vanishing ${\cal H}_i$ remains zero in the preferred frame, are all equivalent. 

Dropping these equivalent assumptions is not only allowed: it is a very interesting possibility. But it does lead to a totally different formalism and we leave its development to a separate paper~\cite{LIVCDM}.


\section{Effective dark matter}


Returning to $S_m$ defined in \eqref{SmprefS}
we start by making statements in coordinates adapted to $\Sigma_t$. Action \eqref{SmprefS} implies the equations of motion:
\begin{align}
    \dot m=&\partial_i(N^i m)\label{dotm}\\
    \dot \phi =&N+N^i\partial_i\phi,\label{dotphieq}
\end{align}
which are an abridged version of \eqref{dotHH} and \eqref{dotHHi},  since 
\begin{align}
    {\cal H}_m&=m\nn\\
    {\cal H}_i^m&=m \, \partial_i\phi\nn
\end{align}
with the additional constraint ${\cal H}^m_i=0$. We see that \eqref{dotm}
states that $S_m$ produces the Lie dragging equation \eqref{LieDrag}, as claimed in Section~\ref{Sec:restor}. In addition, under the assumption that $\Sigma_t$ is geodesic, $N=N(t)$; therefore \eqref{dotphieq} 
implies that the constraint $\partial_i\phi=0$ is preserved by the time evolution, with on-shell identities $n_0=-N=-\dot \phi$ in these coordinates, implying the covariant expression: 
\begin{equation}\label{nmuexp}
    n_\mu=-\partial_\mu \phi. 
\end{equation}
We see that $\phi$, the conjugate variable to $m$, is (minus) the fluid proper time variable for dust matter.


The only other equations of motion affected by $S_m$ are the gravitational equations, in which $S_m$ enters via its implied stress energy tensor. In coordinates adapted to $\Sigma_t$ this has only one non-vanishing component:
\begin{equation}
    T_{00}^m=\frac{-2}{N\sqrt{h}}\frac{\delta S_m}{\delta g^{00}}=
   N^2 \frac{m}{\sqrt{h}},
\end{equation}
and can therefore be written in covariant form as:
\begin{equation}
   T^m_{\mu\nu}
    =\rho_m(x)n_\mu n_\nu,
\end{equation} 
where
\begin{equation}
    \rho_m=\frac{m}{\sqrt{h}}
\end{equation}
is a 4D scalar (i.e. a density of weight zero). This is the stress-energy tensor of a pressureless perfect fluid. Variation of $\bar S_0$ with respect to the metric shows that $T_{\mu\nu}^m$ appears in the right hand side of the gravitational equations as an additional term:
\begin{equation}
    \bar T_{\mu \nu}=T_{\mu\nu}+T^m_{\mu\nu}
\end{equation}
(where $T_{\mu \nu}$ refers to all the forms of matter contained in $S_0$), 
so it has the usual {\it active} gravitational properties.

It is also independently conserved:
\begin{eqnarray}
    \nabla_\mu T^{\mu\nu}_m=0
\end{eqnarray}
as we now prove.
First, $n^\mu$ is by construction geodesic, so: 
\begin{equation}
    h^\alpha_{\; \; \nu} \nabla_\mu T^{\mu\nu}=\rho n^\mu\nabla_\mu n^\alpha= 0.
\end{equation}
Second, the Lie dragging equation \eqref{LieDrag} implies the other component of the conservation equation, (constancy of the dust grains' rest mass), $\nabla_\mu (\rho n^\mu)=0$.
Recalling \eqref{n^mu}, we have:
\begin{equation}
    -n_\nu \nabla_\mu T^{\mu\nu}=\nabla_\mu (\rho n^\mu)=\frac{1}{\sqrt{-g}}(\dot m -(m N^i)_{,i})=0
\end{equation}
that is, the relativistic conservation equation:
\begin{equation}
    \dot\rho +\rho\nabla_\mu n^\mu=0.
\end{equation}

\section{Expression in a general frame}


All of the above was done in coordinates adapted to the preferred $\Sigma_t$, but we may not know this frame,
so it is useful to write the frame-selecting action $S_m$ in generic coordinates.
Let us change notation, replacing $n_\mu\rightarrow u_\mu$ in the previous section, so that $u_\mu$ denotes the normal of the special foliation $\Sigma_t$, releasing $n_\mu$ to stand for the normal to a generic ADM foliation. From the expression for the ``gamma factor'' of a relativistic fluid~\cite{brown}: 
\begin{equation}\label{gamma}
    -n^\mu u_\mu =\sqrt{1+h^{ij} u_i u_j}\equiv \gamma(u_i,h^{ij})
\end{equation}
we can write in a generic frame:
\begin{align}
    u_\mu&=(N^i u_i - N \gamma, u_i)\\
     u^\mu&=\left(\frac{\gamma}{N}, u^i-\frac{N^i\gamma}{N}\right)
\end{align}
where $N$, $N^i$ and $h_{ij}$ are defined in the new foliation and we define:
\begin{equation}
    u^i\equiv h^{ij} u_j
\end{equation}
not to be confused with the 4-D counterpart
$g^{i\mu}u_\mu=h^{ij}u_j-N^i\gamma/N$. Inserting this form of $u_\mu$ in:
\begin{equation}\label{Tmu-u}
   T^m_{\mu\nu}
    =\rho_m(x)u_\mu u_\nu,
\end{equation} 
we get the stress-energy tensor expression in any frame. This provides the effect on the gravitational equations of motion in any coordinates or foliation.

Likewise, the general form of the Hamiltonian and momentum can be read off from \eqref{Tmu-u}, using 
relations: 
\begin{align}
     {\cal H}_m&= \sqrt{h}n^\mu n^\nu T^m_{\mu\nu}\label{HinT}\\
     {\cal H}^m_i&=
     \sqrt{h}n^\mu P^\nu_{\; i} T^m_{\mu\nu}.\label{HiinT}
\end{align}
Recall that, although $\rho_m$ is a 4D scalar, $m(x)$ is a spatial density. If $m$ and $h_S$ are the Hamiltonian and determinant of the spatial metric in the original frame, and $m(u_i)$ and $h$ those in the new one, we have: 
\begin{equation}
    m(u_i)=m\sqrt{\frac{h}{h_S}}.
\end{equation}
Using coordinates adapted to the new frame we have $h_{ij}=g_{ij}+n_i n_j=g_{ij}$, whereas the induced metric in the old (preferred) foliation is $h^{(S)}_{ij}=g_{ij}+u_i u_j$. Their determinants are thus related by: 
\begin{equation}
    h_S=h(1+h^{ij}u_i u_j)
\end{equation}
so that:
\begin{equation}
    m(u_i)=\frac{m}{(1+h^{ij}u_i u_j)^{1/2}}.
\end{equation}
Using \eqref{HinT}-\eqref{HiinT} and \eqref{gamma} we finally have:
\begin{align}
     {\cal H}_m&=m(u_i) (n^\mu u_\mu)^2=m(1+h^{ij}u_i u_j)^{1/2}\label{Hfluid}\\
     {\cal H}^m_i&=
     m(u_i) (n^\mu u_\mu)u_i
     = - mu_i .\label{Hifluid}
\end{align}
Given \eqref{nmuexp}, in a general frame we should replace $\lambda^i\partial_i \phi$ in \eqref{SmprefS} by $\lambda^i(n_i+\partial_i \phi)$. We can also simply ignore this constraint and write the action as
\begin{align}\label{SmgenS}
     S_m&=
     \int dt\, d^3x\, (\dot \phi(x) m(x)
     -N {\cal H}_m -N^i  {\cal H}^m_i )
\end{align}
where it is understood that $u_i$ is shorthand for:
\begin{equation}\label{uidef}
    u_i\equiv -\partial_i\phi.
\end{equation}
This is just the Brown action~\cite{brown} for a relativistic fluid, freezing out the entropy
and hydrodynamical velocity degrees of freedom\footnote{See 3.10 and 3.11 with 3.7 and 3.9 in Ref.~\cite{brown}. See also \cite{brownkuchar}.}. Hence the full legacy of past evolution is just a form of cold dark matter (at least assuming that $\Sigma_t$ is geodesic). However, as we now show, this is a very special type of dark matter. 

\section{Preferred geodesic frames and special dust}
Freezing the entropy and hydrodynamical degrees of freedom makes sense and completes our discussion of symmetries and degrees of freedom. We have found that theories with global interactions have 3 local gravitational degrees of freedom, having lost one local constraint (the Hamiltonian constraint). After global interactions switch off the extra degree of freedom can be exported into a matter-like component, since 
we can separate the action into an action $S_0$ with full diffeomorphism invariance restored (and thus the usual two gravitons), and an effective fluid action absorbing the extra degree of freedom (which therefore is not a propagating ``graviton''). 

This is necessarily a very special fluid. 
A generic perfect fluid has 5 degrees of freedom: the number density, the entropy density and the freedom the fluid has to move in 3 spatial directions. These correspond to 5 unconstrained pairs of variables (see e.g.~\cite{brown}). Since we are absorbing only one degree of freedom into this effective fluid, the fluid has to have 4 degrees of freedom ``frozen''. The entropy degree of freedom is erased by the dust equation of state $p=0$. The hydrodynamical degrees of freedom are frozen 
by the assumption that $u_\mu$ is not dynamical, but is a pre-given preferred frame (albeit geodesic, according to the assumptions made here). The counting therefore matches. 

Such ``freezing'' is better formalized by showing its equivalence to imposing first class constraints on a generic fluid. The dust condition, like any other equation of state, can be seen as a first class constraint on a more generic fluid since dust remains (or can remain) dust under time evolution. In addition we have to impose 3 conditions on the most general dust action, $S_d$. Such action can be written (see~\cite{brown,brownkuchar}): 
\begin{align}\label{Sdust}
     S_{d}&=
     \int dt\, d^3x\, (\dot \phi m+ \dot\alpha^A (m\beta_A)
     -N {\cal H}_m -N^i  {\cal H}^m_i )
\end{align}
with ${\cal H}_m$ and  ${\cal H}^m_i$ still given by \eqref{Hfluid} and \eqref{Hifluid}, but with \eqref{uidef} replaced by:
\begin{equation}
    u_i=-(\partial_i\phi +\beta_A\partial_i\alpha^A).
\end{equation}
Here $\alpha^A$ and $\beta_A$ ($A=1,2,3$) are 6 space-time scalars containing the hydrodynamical degrees of freedom. The $\alpha^A$ are the Lagrangian or ``comoving'' coordinates of the fluid, and $m\beta_A$ are their conjugate momenta. They evolve under~\cite{brown,brownkuchar}:
\begin{align}
    \dot\alpha^A&=\{\alpha^A,\mathbf{H}\}=-\frac{N}{\gamma}u^i\partial_i\alpha^A +N^i\partial_i \alpha^A\label{dotalpha}\\
      (m\beta_A)^.&=\{m\beta_A,\mathbf{H}\}=-\partial_i (Nm\beta_Au^i)+\partial_i(N^im\beta_A).\label{dotbeta}
\end{align}
Hence 
\begin{equation}
    \beta_A=0
\end{equation}
is the required set of 3  constraints\footnote{This choice breaks the fluid symmetries described in~\cite{brown}, but the matter will not be relevant here.}.
If $\beta_A=0$ (and so also $\partial_i\beta_A=0$) the evolution equation \eqref{dotbeta} implies $\dot \beta_A=0$, so the constraints are first class and do remove or ``freeze'' the 3 hydrodynamical degrees of freedom of the fluid.  The other evolution equation \eqref{dotalpha} is just  the transformation law between Eulerian and Lagrangian coordinates. The $\alpha^A$ and $\beta_A$ do not contribute to $u^\mu$, or to the extended Hamiltonian. They drop out of the remaining dust equations of motion:
\begin{align}
    \dot\phi&=\{\phi,\mathbf{H}\}=N\gamma +N^i\partial_i\phi\\
    \dot m&=\{m,\mathbf{H}\}=-\partial_i (Nm u^i)+\partial_i(N^im)
\end{align}
which are just \eqref{dotm} and \eqref{dotphieq} in a generic frame (i.e. not in the rest frame $\Sigma_t$ or dust). 

\section{``Painted on'' dark matter and testability}\label{Sec:test}

Hence, even with the geodesic assumption, a special frame remains in place. The effective fluid absorbing the extra degree of freedom is not (and could not be, as we know just from ``counting'')  the most general dust fluid, for which the introduction of a foliation is just a convenient mathematical device with no physical significance. The rest frame of the equivalent dust fluid is pre-fixed, among all possible geodesic frames, so we still have what in~\cite{MachianCDM} was labelled ``painted on'' dark matter.

But how could we distinguish it from conventional dark matter? Unlike the case in~\cite{LIVCDM}, where a non-geodesic $\Sigma_t$  leaves dramatic new imprints, the distinction here is subtle. Apart from the fact that vortical motion could never be assigned to our effective fluid\footnote{The flow is a gradient flow, a requirement of Frobenius theorem applied to $\Sigma_t$.}, and barring pre-knowledge of the preferred frame, we can only distinguish our effective matter from conventional dark matter by the preferred frame being something that for normal dark matter would be ``outlandish'', in the sense that it would require the initial conditions having an uncanny pre-cognition of what would come in the future. We give a simple example, which may have phenomenological significance in an improved form.


Consider the Lemaitre-Tolman-Bondi (LTB) formalism~\cite{exactslns,enqvist,Goncalves}, starting from metric:
\begin{equation}\label{pre-LTB}
    ds^2=-dt^2+X^2(r,t) dr^2+A^2(r,t)d\Omega^2,
\end{equation}
which only assumes spherical symmetry. This is a geodesic frame, since $N=1$. Although $N^i=0$ there is a momentum constraint (obtained by varying $N^i$ around zero). If this frame is the preferred frame $\Sigma_t$, then ${\cal H}_i^m=0$, and assuming all other forms of dust matter are comoving with it (or vanish) in the region and epoch of interest, the momentum constraint reads:
\begin{align}
    \dot{A}'-A'\frac{\dot{X}}{X}&=0\label{mom}
\end{align}
(where $^.\equiv \partial/\partial t$ and $'\equiv \partial/\partial r$). This equation has a first integral which places the metric in the form:
\begin{equation}\label{LTB}
    ds^2=-dt^2+\frac{A'^2}{1-K(r)}dr^2+A^2d\Omega^2.
\end{equation}
(We stress that this step would not be valid if ${\cal H}_i^m\neq 0$.) 
The remaining equations then amount to the local Hamiltonian constraint:
\begin{eqnarray}
  \frac{\dot A^2}{A^2}&=&\frac{F(r)}{A^3}-\frac{K(r)}{A^2}
\end{eqnarray}
with:
\begin{equation}
   F'=8\pi G\rho A'A^2.
\end{equation}
The FRW metric and equations can be recognized within the LTB series by setting $A=a(t)r$, $K=kr^2$ (with $a$ the expansion factor and $k=0,\pm 1$ the normalized intrinsic  curvature) as well as:
\begin{equation}\label{FrFRW}
     F(r)=\frac{8\pi G}{3}r^3 (\rho_{tot} a^3) = \frac{8\pi G}{3}r^3 m_{tot}
\end{equation}
(where $m_{tot}=\rho_{tot} a^3$ refers to the constant of motion for the total pressureless component). This suggests that the $t=$const. leaves in these coordinates could be the preferred frame $\Sigma_t$ inherited from a cosmological origin\footnote{The label $r$ is also fixed by the form of \eqref{FrFRW}. We could call this the Hubble gauge, noting that in general LTB has a residual freedom of redefinition of $r$.}. This is obviously true in the FRW approximation of~\cite{MachianCDM} (and of~\cite{shinji} in the context of Horava-Lifshitz theory~\cite{HL}), but the LTB framework allows us to consider inhomogeneous solutions. These could include primordial inhomogeneities in $m(x,t_0)$, but more simply result from situations where a primordial homogeneous $m(x,t_0)=m(t_0)$ reacts to the condensation of structures made from other forms of matter.

Let us therefore assume that the preferred frame is the LTB frame with $K=0$ matching a primordial and a present asymptotic FRW Universe. This frame remains present even when primordial perturbations are added on, or large scale structures condense at late times. As a test tube for the implications of this statement, let other pressureless matter form a central black hole with Schwarzchild radius $R_s=2GM$ creating a spherical hole in an otherwise $k=0$ FRW Universe. 
If the LTB frame is $\Sigma_t$, obviously our effective DM cannot condense, since $N^i=0$ makes Lie dragging \eqref{LieDrag} trivial. Hence the effective matter will continue to expand even though other forms of matter condensed, creating the first anomaly.

Eq.~\eqref{LieDrag} implies $\dot m=0$ where   $m$ is a spatial density, which is uniform in Cartesians ($m=m_0$).  By evaluating the relevant factors of $h$ (first from Cartesians to polars, then to full LTB) we find that\footnote{We could also derive this expressions from the LTB formula~\cite{enqvist,Goncalves}:
\begin{equation}
  \rho_m = \frac{F'_m(r)}{8\pi G A' A^2}
\end{equation}
using \eqref{FrFRW}  and noting that $F(r)$ is additive.}:
\begin{equation}\label{rhom}
    \rho_m(r,t)=\frac{m_0 r^2}{A^2|A'|}.
\end{equation}
In the inner region where the black hole dominates (or assuming $\Omega_m\ll 1$, so that the escavated mass $M$ always dominates~\cite{Stephani}) we have the vacuum solution~\cite{exactslns}:
\begin{equation}
    A(r,t)=\left(\frac{9R_s}{4}\right)^{1/3}(t+r-r_0)^{2/3}\label{sollast}
\end{equation}
in the geodesic outgoing coordinates (with $A'>0$) associated with $\Sigma_t$. We have adjusted constants so that the FRW solution is matched at $r=r_0$, setting $t=0$ for the Big Bang time. The other matching condition equates $M$ with the escavated mass taken out of the FRW model to create the vacuole (e.g.~\cite{Stephani}), or:
\begin{equation}
    \frac{8\pi G}{3}\rho_c a^3 r_0^3=R_s\implies
    H^2 A_0^3 = R_s
\end{equation}
(where $\rho_c$ is the critical cosmic density, H is the Hubble parameter, and $A_0=ar_0$). 

Eqs.~\eqref{rhom} and \eqref{sollast} imply:
\begin{equation}
    \rho_m=\frac{m_0r^2}{A^{3/2}\sqrt{R_s}}=\frac{2m_0r^2}{3 R_s(t+r-r_0)}
\end{equation}
and since this is a scalar we know the density in any coordinates (modulo the fact that different coordinate charts may cover different regions). In particular, we can transform to Schwarzchild coordinates
$ds^2=-dT^2(1-R_s/A)+ dA^2/(1-R_s/A)+A^2d\Omega^2 $, noting that $\{t,r\}$ for this solution are the Lemaitre coordinates~\cite{exactslns} (with $A'=\sqrt{R_s/A}$)\footnote{We recall the transformation $dt=dT-dA/(\sqrt{x}(1-1/x))$ and $dr=-dT+\sqrt{x}dA/(1-1/x)$, where $x=A/R_s$. }. In fact, all we need is:
\begin{align}
    r=-T+R_s\int ^{A/R_s}\frac{dx\, \sqrt{x}}{1-\frac{1}{x}}
\end{align}
(we spare the reader the explicit expression).
The integral is logarithmically divergent at the black hole horizon, $A=R_s$, which is the boundary of the patch covered by the LTB outgoing coordinates; this leads to a divergence in $\rho_m$. The effects on black hole formation could be interesting. 

In the Newtonian region:
\begin{equation}
    r\approx -T+r_0 +\frac{2}{3\sqrt{R_s}}A^{3/2}
\end{equation}
where we adjusted the integration constants so that this matches \eqref{sollast} with $T\approx t$. Fixing $r$, we can recognize the Newtonian trajectory of particles with exact escape speed, starting from delayed ``Big Bangs''\footnote{In the conventional sense of times for which $A=0$ and $\dot A=\infty$ extrapolating back in time, {\it erroneously} using the Newtonian picture all the way back.}. Unlike with a Newtonian Hubble flow, where the explosion started at the same time for all the shells and each shell is slowed down by the gravity of all the inner shells, in our case the explosion started later for the inner shells, and all shells feel the gravity due to the same dominant central mass $M$. Specifically, we see a flow satisfying Newton's laws for a central mass $M$:
\begin{equation}
    T-T_{BB}(r)=\frac{2}{3\sqrt{R_s}}A^{3/2}
\end{equation}
with $T_{BB}=r_0-r$. The density profile in the Newtonian region is:
\begin{align}\label{profile}
    \rho_m(A,T) &=\frac{m_0}{A^{3/2}\sqrt{R_s}}
\left(-T+r_0 +\frac{2}{3\sqrt{R_s}}A^{3/2}\right)^2
\end{align}
with 
\begin{align}
&\frac{2}{3} \frac{A^{3/2}}{R_S^{1/2}}<T<r_0+\frac{2}{3} A^{3/2}/R_S^{1/2}\nn
\end{align}
so no longer do we have a homogeneous, time dependent profile as in the case of a Hubble flow. Instead we can think of this profile as a halo with roughly:
\begin{equation}
     \rho_m(A,T) \approx \frac{m_0r_0^2}{A^{3/2}\sqrt{R_s}}
\end{equation}
(for $A\gg R_s$), that is with a roughly static profile $1/A^{3/2}$ for a large period of its 
life time $\Delta T\approx r_0$. Eq. \eqref{profile} complements the details of this simplified picture. 



This would be a very strange kind of ``halo'' if made of normal matter. It would consist of dust particles moving outwards along geodesics which are on the border between bounded and unbounded, each with a delayed starting time according to a well defined rule. This would happen even when other types of pressureless matter collapsed to produce a structure (presumably due to being in a bounded Hubble flow from the start). Such ``halo'' would have a density profile neither matching the Hubble flow it came from, nor the current condensed system into each it is embedded. 
Barring conspiracy initial conditions, 
where would such a structure come from if not from the presence of a preferred cosmological geodesic frame?

This is just an example, and more elaborate models can be built, as we now sketch. Depending on the size of the vacuole with regards to the Hubble length, this structure might more closely resemble an ejected corona than a halo. We could also consider non-homogenous initial $m(x)$, or situations where $\rho_m$ is not subdominant inside condensed structures. Could this help explain galaxy rotation curves, bypassing the shortcomings of more conventional dark matter, such as the cusp-core problem? Elsewhere we will also investigate how this effect could resolve the problem of caustic formation. 


\section{Conclusions}

In this paper we investigated in a general setting the effects of breaking full diffeomorphism invariance down to spatial diffeomorphism invariance on a foliation, and then restoring it. 
A new local degree of freedom is created by such partial symmetry breaking, which persists even after full symmetry is restored. If the foliation breaking the symmetry is geodesic by construction, this degree of freedom is equivalent to a dark matter component, which does not need to be homogeneous. However, only one out of its four local degrees of freedom is available. Phenomenological implications were discussed. 

Our general setting involved the action upon local physics of {\it global} variables, that is variables defined on the whole leaves of the foliation. This is to be distinguished from perturbatively non-local interactions, such as those with higher order derivatives. We stress that we are envisaging the effects of variables that exist globally on $\Sigma_t$, rather than the more traditional setting of terms in the action that depend non-locally on local variables\footnote{Indeed such perturbatively non-local terms could be bundled into what we called local action, which should perhaps more rigorously be dubbed ``non-global''.}. Global interactions break locality maximally and introduce a preferred frame or foliation. They have a ``Machian'' flavour, within a meaning backtracking to before the advent of General Relativity, as pointed out in~\cite{MachianCDM}. Our considerations have nothing to do with inertia, but they do draw on Mach's approach~\cite{MachBook,MachReview} in that we are entertaining direct instantaneous action of the whole Universe on its parts. 

Several types of theory fall within this structure. Our original motivation was work on variability of the laws of physics (\cite{evol,BHevol} and more specifically \cite{MachianCDM}), itself grounded on unimodular-like theories~\cite{unimod,unimod1,UnimodLee1,alan,daughton,sorkin1,sorkin2,Bombelli,UnimodLee2,JoaoLetter,JoaoPaper,pad,pad1,lomb,vikman,vikman1,vikaxion,JoaoLetter,JoaoPaper}. In this context the global variables are the (pseudo-)constants of Nature and their canonically dual times, the  latter defined as in~\cite{unimod} (and {\it not} on the past light cone of an observer, as often done in the context of causal set theory).  If time is evaluated on space-like hypersurfaces, any time dependence of the local Hamiltonian gives physical relevance to the foliation  (just as, conversely, time independence and on-shell space-time constancy of constants erases its physical relevance~\cite{unimod}). 
Despite this declared motivation, the results derived here are likely applicable with minimal modification to other theories with the same structure.  We mention Horava-Lifshitz theory  (e.g.~\cite{HL,shinji}), some approaches to time in Quantum Gravity (e.g. Sec. 4.3.2 of~\cite{Isham} and references therein; also~\cite{brownkuchar,viqar}) and more generally the issue of gauge fixing in Quantum Gravity (e.g.~\cite{StueckelDM,Barvinsky}) and how it may introduce a conceptual preferred frame. Theories based on Dirac's non-linear electrodynamics may also have a preferred foliation~\cite{DBI} leading to the same effects, possibly compounded by the fact that their preferred length scale could itself be turned into a global variable\footnote{We are grateful to the anonymous referee for this suggestion.}.


In addition to these ramifications, 
our results motivate much further work. The existence of a preferred frame introduces an element of cosmic {\it precognition} common to all theories with a global structure (e.g.~\cite{pad,pad1,padilla}\footnote{The term ``pre-cognition'' in relation to the cosmological constant was first used by Coleman~\cite{Coleman}.}). As we saw in Section~\ref{Sec:test}, this can be used to distinguish the associated effective matter from conventional dark matter, in the sense that the latter would need to have its initial conditions incredibly fine tuned to reproduce what a preferred geodesic frame naturally gives to our effective dust, with three of its four degrees of freedom frozen. We gave an example in Section~\ref{Sec:test}, but could a more elaborate model serve as partial explanation, for example, of the galaxy rotation curves? Astrophysical issues such as the flattening of the galaxy rotation curves and the Tully-Fisher relation might then be truly cosmological initial condition problems. 

More dramatically, the preferred $\Sigma_t$ might not be geodesic as explored in the sequel to this paper~\cite{LIVCDM}.

\section{Acknowledgments}
We thank Niayesh Afshordi, Stephon Alexander, Paolo Bassani, Claudia de Rham,  Shinji Mukohyama and Tom Zlosnik for discussions. 
This work was partly supported by the STFC Consolidated Grants ST/T000791/1 and ST/X00575/1.

\end{document}